# HIDDENdb: Co-dependency database reveals a plethora of genetic and protein interactions

**Iresha De Silva**[1,2], **Shantha Pathma Bandu**[1], **Rune T. Kidmose**[1], **Genona T. Maseras**[1,2], **Thomas Bataillon**[2], **Xavier Bofill-De Ros**[1]*

[1]Department of Molecular Biology and Genetics, Aarhus University, Denmark
[2]Section for Bioinformatics Research and Computational Biology (BiRC), Aarhus University, Denmark
*Correspondence should be addressed to X.B.D-R. (xbofill@mbg.au.dk)

**Genetic interactions and protein co-dependencies shape cellular fitness, buffering capacity, and disease vulnerability. However, systematic integration of co-dependency relationships across heterogeneous datasets remains limited. Here, we present HIDDENdb (Harnessing Intelligent Data Discovery to Explore Gene Networks), a comprehensive database that captures genetic and protein co-dependencies inferred from large-scale perturbation screens, multi-omics datasets, and curated interaction repositories. HIDDENdb integrates genome-wide loss-of-function screens (CRISPR and shRNA) with other unbiased resources (BIOGRID-ORCS and GWAS) to construct a map of co-dependency relationships across diverse biological contexts. Using robust statistical modeling and network inference approaches, we identify modules of genes and proteins exhibiting shared dependency patterns across cell lines. Notably, top-ranked gene-gene co-dependency pairs are enriched for high-confidence AlphaFold-predicted protein–protein interfaces, suggesting that a subset of inferred functional relationships may reflect underlying structural interactions. Importantly, the database enables users to explore co-dependency networks interactively. HIDDENdb is freely accessible through a web-based interactive interface at https://bofillderoslab.shinyapps.io/hiddendb/.**

## Introduction

Despite extensive progress in functional genomics, a considerable proportion of human protein-coding genes remain poorly characterized, lacking clear functional annotation, pathway assignment, or mechanistic insight [1,2]. This gap in molecular biology limits our capacity to systematically interpret genetic variation, identify novel disease drivers or therapeutic targets. Historically, functional knowledge has accumulated through targeted studies of individual genes, an approach that has been highly informative but inherently biased toward well-studied pathways and phenotypes [3].

Genome-wide and unbiased methodologies now enables systematic interrogation of protein-coding gene function at scale. High-throughput perturbation screens, population genetics resources, interaction datasets, and integrative multi-omics analyses provide complementary perspectives on gene function without requiring prior assumptions. Leveraging these approaches is therefore essential to uncover hidden functional relationships, define genetic dependencies, and assign roles to understudied protein-coding genes. To support this effort, we developed HIDDENdb, a resource that integrates genome-wide perturbation data, genetic wide association studies, and interaction repositories into a unified co-dependency framework. HIDDENdb provides a systematic platform to explore genes with shared dependency patterns across available datasets. We hope this can aid functional annotation and suggest candidate genetic relationships for further experimental validation.

## Results

HIDDENdb is organized into multiple tabs that facilitate exploration of the datasets through summary tables and interactive visualizations. Genetic interactions for individual genes can be queried using dynamic plots that display the top co-dependent partners across all chromosomes, together with the direction and strength of the interaction within each dataset (Achilles, Sanger, or shRNA) *(Fig. 1A)*. In these visualizations, dot size reflects the number of additional shared interactors between the query gene and the candidate partner, providing an indication of network overlap. A color scheme further denotes whether interactions have prior supporting evidence in repositories of physical or spatial interactions, including OpenCell [10] or BioGRID v4.4 [7]. Simultaneous cross-comparison between the two major CRISPR–Cas9 datasets facilitates the identification of robust gene–gene interactions that may be less apparent when analyzing each dataset independently *(Fig. 1B)*. In addition, HIDDENdb enables visualization of the first-degree neighbors of top co-dependent genes, allowing users to explore local network structure and to generate hypotheses regarding higher-order molecular complexes or pathway organization *(Fig. 1C)*.

Beyond visualization of pairwise interactions, HIDDENdb provides several features that support exploratory and hypothesis-driven analyses. Users can filter interactions by strength, reproducibility across datasets, or supporting evidence from external resources, enabling prioritization of high-confidence candidates. Users can export of interaction tables, all visualised plots and network structures for downstream analyses, facilitating integration with external bioinformatics workflows. Source code is available on GitHub, and comprehensive documentation and usage guidelines are provided through the project Wiki (https://github.com/Bofill-De-Ros-Lab/HIDDENdb/wiki).

### Application of HIDDENdb to understudied genes

As an illustrative example, we examined ZCCHC7 (zinc finger CCHC domain-containing protein 7). Querying ZCCHC7 in HIDDENdb identifies TENT4B (also known as PAPD5 or TUT3) as a prominent co-dependent partner across datasets, with a Z-score of 9.8 **(Fig. 1A,B)**. This finding is consistent with prior evidence linking ZCCHC7 and TENT4B to the TRAMP-like complexes [11,12], in which the non-canonical poly(A) polymerase operates together with zinc-finger RNA-binding proteins to promote RNA decay. Interestingly, other common genetic interactions between both genes uncover CD1, PINX1 and DDX21 (Fig. 1C), point towards their known roles in the exosome [13], telomere-associated processes [14,15], as well as more broad RNA metabolism [16], respectively. As a second example, we examined RBM48 (RNA Binding Motif Protein 48), a relatively understudied gene according to the Unknome database (standard knownness score: 0.0) [3]. Among the highest-scoring co-dependent partners, we identified ARMC7 and SCNM1, both previously implicated as components of the minor spliceosome [17] **(Fig. S1A,B)**. Consistent with this, DDX59 has recently been implicated in minor exon splicing [18] **(Fig. S1C,D)**.

### Enrichment of structural interactions among high-confidence co-dependencies

Given the frequent observation that high-confidence genetic interactions correspond to known molecular complexes, we sought to systematically evaluate the extent to which strong co-dependency signals might reflect proven direct physical interactions. To this end, we selected 6,000 gene pairs reported in HIDDENdb and stratified them into four tiers (Q1–Q4) based on their Z-scores. Structural interaction likelihood was assessed using AlphaFold3-derived interface predicted TM (ipTM) scores [19]. Comparing ipTM distributions revealed a clear enrichment of high-confidence structural predictions among top-ranked co-dependencies **(Fig. 1D)**. In the highest tier (Q1), 6.73% of gene pairs exhibited ipTM scores >0.8, and 17.21% exceeded 0.6. In contrast, among the lowest tier (Q4), only 1.84% of pairs surpassed 0.8 and 7.89% exceeded 0.6. These results indicate that strong co-dependency signals are significantly enriched for predicted physical protein–protein interactions, although most genetic interactions may reflect indirect or pathway-level functional relationships.

## Discussion

HIDDENdb provides a systematic framework to explore genetic interactions and co-dependencies, particularly for poorly characterized protein-coding genes. By integrating independent perturbation datasets and complementary annotation layers, the platform enables the identification of co-dependency signals that frequently align with known molecular complexes and pathways. As illustrated in the case studies presented here, querying understudied genes readily uncovers biologically coherent interaction partners, facilitating hypothesis generation and functional inference in the absence of extensive prior literature. In this context, HIDDENdb serves as a scalable tool to accelerate functional study and to prioritize candidate molecular associations for downstream validation.

Interpretation of co-dependency signals requires caution. For instance, TENT4A (PAPD7/TUT5), a close paralog of TENT4B often considered functionally interchangeable [20], showed no comparable co-dependency signals, and ZCCHC14 likewise displayed no interaction with TENT4A [21,22]. These findings suggest that co-dependency patterns can reveal specificity in molecular complex organization that may not be evident from biochemical assays alone. Conversely, absence of co-dependency does not preclude functional overlap, as buffering, context dependence, or compensatory mechanisms may obscure relationships in pooled screens.

Systematic biases should also be considered. Mitochondrial genes frequently display strong co-dependency signals, reflecting the tightly coordinated function of oxidative phosphorylation and mitochondrial complexes, but potentially inflating connectivity within organelle-restricted modules. In addition, genes located in close chromosomal proximity show correlated dependency profiles in CRISPR screens due to copy-number effects or collateral damage from double-strand breaks. Such regional effects can generate apparent co-dependencies that reflect shared genomic context rather than direct functional interaction.

## References

1. Edwards, A.M., Isserlin, R., Bader, G.D., Frye, S.V., Willson, T.M., and Yu, F.H. (2011). Too many roads not taken. Nature 470, 163–165.

2. Stoeger, T., Gerlach, M., Morimoto, R.I., and Nunes Amaral, L.A. (2018). Large-scale investigation of the reasons why potentially important genes are ignored. PLoS Biol. 16, e2006643.

3. Rocha, J.J., Jayaram, S.A., Stevens, T.J., Muschalik, N., Shah, R.D., Emran, S., Robles, C., Freeman, M., and Munro, S. (2023). Functional unknomics: Systematic screening of conserved genes of unknown function. PLoS Biol. 21, e3002222.

4. Dempster, J.M., Pacini, C., Pantel, S., Behan, F.M., Green, T., Krill-Burger, J., Beaver, C.M., Younger, S.T., Zhivich, V., Najgebauer, H., et al. (2019). Agreement between two large pan-cancer CRISPR-Cas9 gene dependency data sets. Nat. Commun. 10, 5817.

5. Pacini, C., Dempster, J.M., Boyle, I., Gonçalves, E., Najgebauer, H., Karakoc, E., van der Meer, D., Barthorpe, A., Lightfoot, H., Jaaks, P., et al. (2021). Integrated cross-study datasets of genetic dependencies in cancer. Nat. Commun. 12, 1661.

6. Tsherniak, A., Vazquez, F., Montgomery, P.G., Weir, B.A., Kryukov, G., Cowley, G.S., Gill, S., Harrington, W.F., Pantel, S., Krill-Burger, J.M., et al. (2017). Defining a cancer dependency map. Cell 170, 564-576.e16.

7. Stark, C., Breitkreutz, B.-J., Reguly, T., Boucher, L., Breitkreutz, A., and Tyers, M. (2006). BioGRID: a general repository for interaction datasets. Nucleic Acids Res. 34, D535-9.

8. Cerezo, M., Sollis, E., Ji, Y., Lewis, E., Abid, A., Bircan, K.O., Hall, P., Hayhurst, J., John, S., Mosaku, A., et al. (2025). The NHGRI-EBI GWAS Catalog: standards for reusability, sustainability and diversity. Nucleic Acids Res. 53, D998–D1005.

9. Rath, S., Sharma, R., Gupta, R., Ast, T., Chan, C., Durham, T.J., Goodman, R.P., Grabarek, Z., Haas, M.E., Hung, W.H.W., et al. (2021). MitoCarta3.0: an updated mitochondrial proteome now with sub-organelle localization and pathway annotations. Nucleic Acids Res. 49, D1541–D1547.

10. Cho, N.H., Cheveralls, K.C., Brunner, A.-D., Kim, K., Michaelis, A.C., Raghavan, P., Kobayashi, H., Savy, L., Li, J.Y., Canaj, H., et al. (2022). OpenCell: Endogenous tagging for the cartography of human cellular organization. Science 375, eabi6983.

11. Sudo, H., Nozaki, A., Uno, H., Ishida, Y.-I., and Nagahama, M. (2016). Interaction properties of human TRAMP-like proteins and their role in pre-rRNA 5′ETS turnover. FEBS Lett. 590, 2963–2972.

12. Fasken, M.B., Leung, S.W., Banerjee, A., Kodani, M.O., Chavez, R., Bowman, E.A., Purohit, M.K., Rubinson, M.E., Rubinson, E.H., and Corbett, A.H. (2011). Air1 zinc knuckles 4 and 5 and a conserved IWRXY motif are critical for the function and integrity of the Trf4/5-Air1/2-Mtr4 polyadenylation (TRAMP) RNA quality control complex. J. Biol. Chem. 286, 37429–37445.

13. Lykke-Andersen, S., Tomecki, R., Jensen, T.H., and Dziembowski, A. (2011). The eukaryotic RNA exosome: same scaffold but variable catalytic subunits. RNA Biol. 8, 61–66.

14. Roake, C.M., and Artandi, S.E. (2020). Regulation of human telomerase in homeostasis and disease. Nat. Rev. Mol. Cell Biol. 21, 384–397.

15. Soohoo, C.Y., Shi, R., Lee, T.H., Huang, P., Lu, K.P., and Zhou, X.Z. (2011). Telomerase inhibitor PinX1 provides a link between TRF1 and telomerase to prevent telomere elongation. J. Biol. Chem. 286, 3894–3906.

16. Calo, E., Flynn, R.A., Martin, L., Spitale, R.C., Chang, H.Y., and Wysocka, J. (2015). RNA helicase DDX21 coordinates transcription and ribosomal RNA processing. Nature 518, 249–253.

17. Bai, R., Wan, R., Wang, L., Xu, K., Zhang, Q., Lei, J., and Shi, Y. (2021). Structure of the activated human minor spliceosome. Science 371.

18. Che, R., Panah, M., Mirani, B., Knowles, K., Ostapovich, A., Majumdar, D., Chen, X., DeSimone, J., White, W., Noonan, M., et al. (2025). Identification of human pathways acting on nuclear non-coding RNAs using the Mirror forward genetic approach. Nat. Commun. 16, 4741.

19. Abramson, J., Adler, J., Dunger, J., Evans, R., Green, T., Pritzel, A., Ronneberger, O., Willmore, L., Ballard, A.J., Bambrick, J., et al. (2024). Accurate structure prediction of biomolecular interactions with AlphaFold 3. Nature 630, 493–500.

20. Yu, S., and Kim, V.N. (2020). A tale of non-canonical tails: gene regulation by post-transcriptional RNA tailing. Nat. Rev. Mol. Cell Biol. 21, 542–556.

21. Kim, D., Lee, Y.-S., Jung, S.-J., Yeo, J., Seo, J.J., Lee, Y.-Y., Lim, J., Chang, H., Song, J., Yang, J., et al. (2020). Viral hijacking of the TENT4-ZCCHC14 complex protects viral RNAs via mixed tailing. Nat. Struct. Mol. Biol. 27, 581–588.

22. Li, Y., Misumi, I., Shiota, T., Sun, L., Lenarcic, E.M., Kim, H., Shirasaki, T., Hertel-Wulff, A., Tibbs, T., Mitchell, J.E., et al. (2022). The ZCCHC14/TENT4 complex is required for hepatitis A virus RNA synthesis. Proc Natl Acad Sci USA 119, e2204511119.

## Acknowledgements

The authors declare that this study received funding from the Lundbeck Foundation (R368-2021-428), and Direktør Michael Hermann Nielsens mindelegat (2024-0093141). IDS was supported by Ranganatha Rajagopal Scholarship. The computational resources used in this project were carried out on the EMCC cluster (https://emcc.au.dk). We gratefully acknowledge EMCC for providing the computational resources and support that made these research results possible.

## Competing interest statement

None declared.

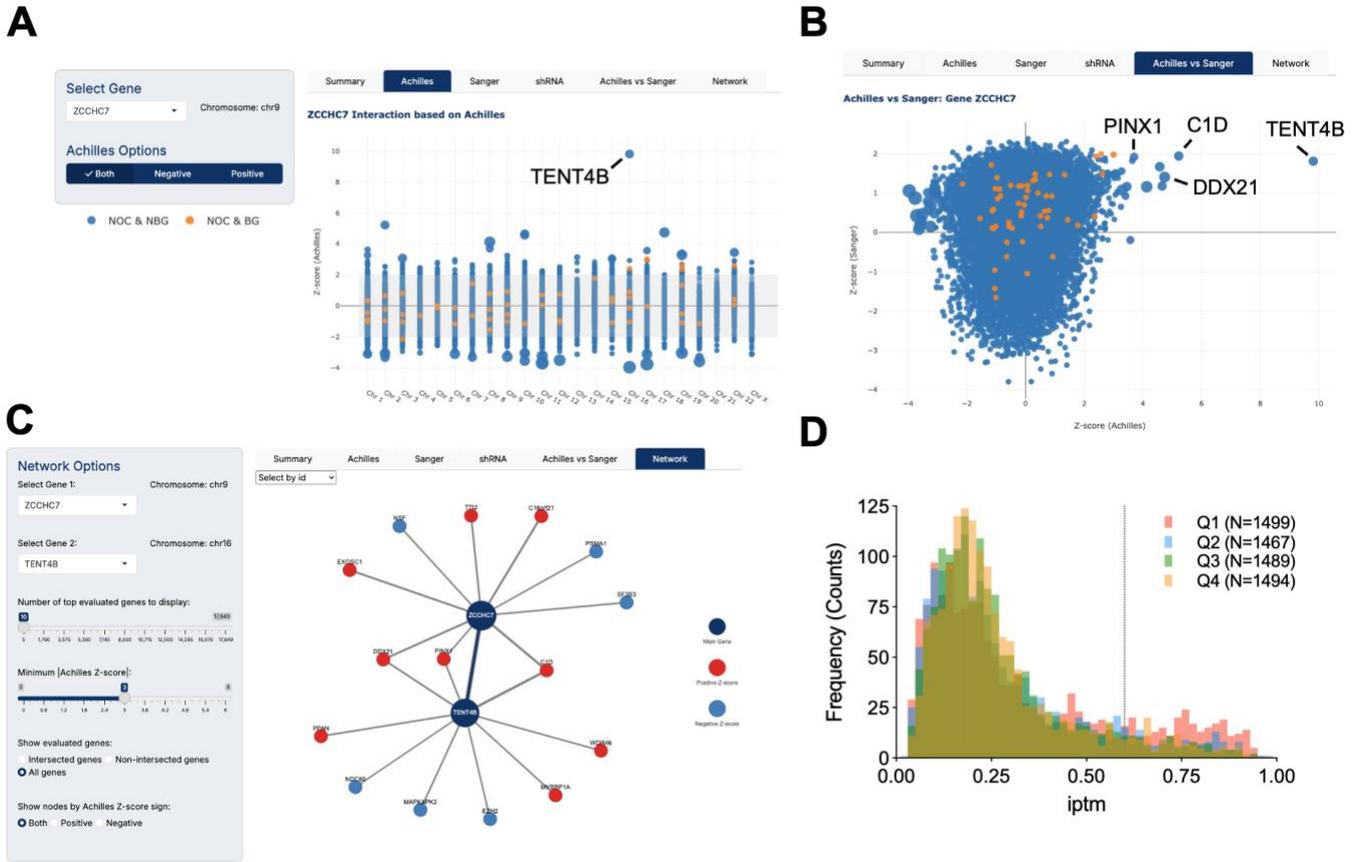

**Figure 1: Visualization and structural validation of gene co-dependencies in HIDDENdb. (A)** Example output of the gene query interface showing chromosome-wide distribution of co-dependency Z-scores for ZCCHC7 in the Achilles dataset. Each point represents a gene, with highlighted interactors indicating top candidates. **(B)** Cross-dataset comparison of dependency scores for ZCCHC7 between Achilles and Sanger CRISPR screens. Genes in the upper-right quadrant represent concordant positive co-dependencies across datasets, highlighting robust interactions such as TENT4B, PINX1, C1D, and DDX21. **(C)** Network visualization of first-degree neighbors for top co-dependent genes, illustrating local interaction structure and shared partners. Node color denotes direction of dependency effect, and edges represent significant co-dependency relationships. **(D)** Distribution of AlphaFold3 interface predicted TM (ipTM) scores for gene pairs stratified by co-dependency strength (Q1–Q4). High-confidence co-dependencies (Q1) show enrichment for higher ipTM values, indicating an increased likelihood of direct physical interaction.

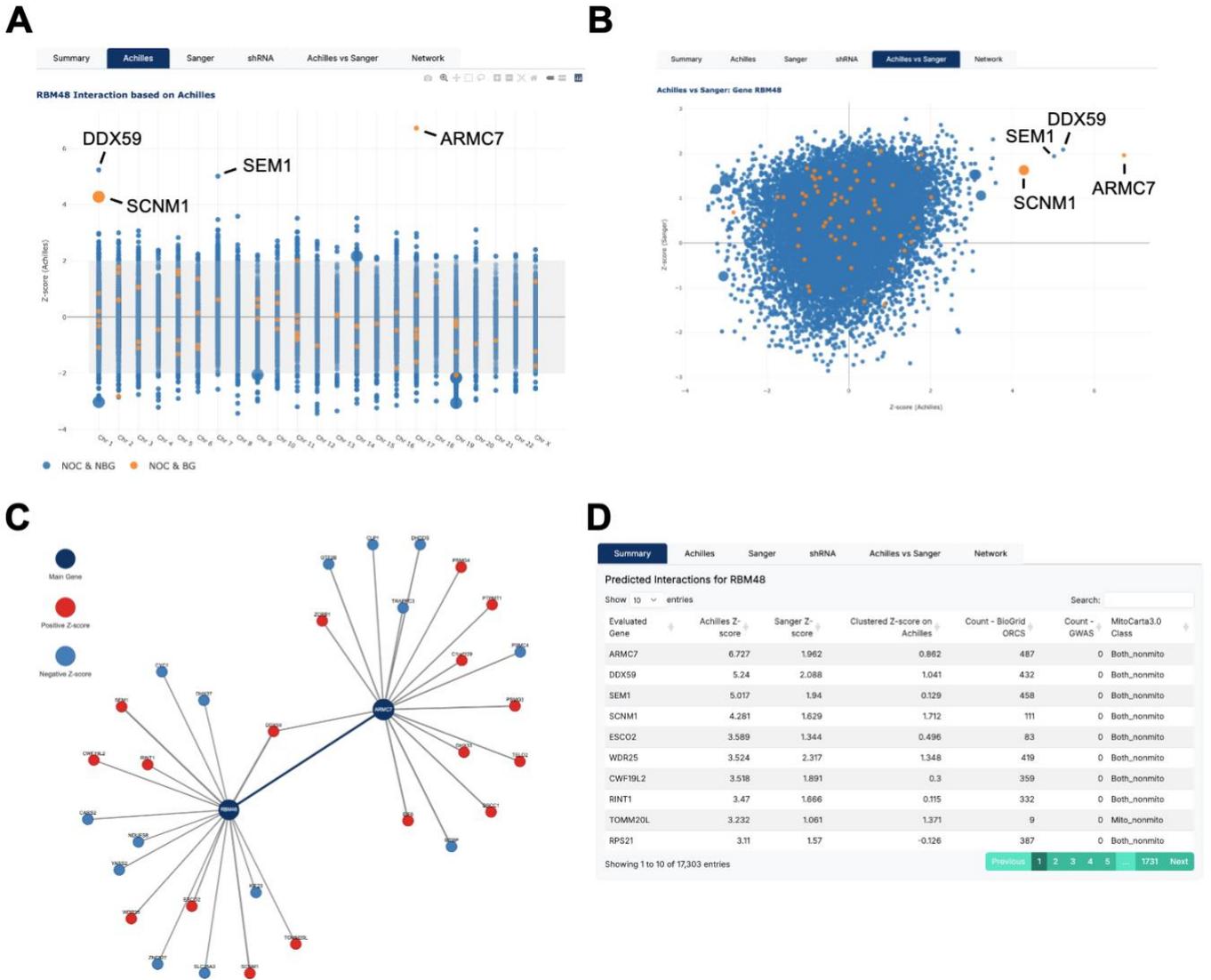

**Figure S1: Identification of RBM48 co-dependencies using HIDDENdb. (A)** Chromosome-wide visualization of co-dependency Z-scores for RBM48 in the Achilles dataset. Each point represents a gene, with highlighted candidates indicating top co-dependent partners, including DDX59, SCNM1, SEM1, and ARMC7. **(B)** Cross-dataset comparison of RBM48 dependency scores between Achilles and Sanger CRISPR screens. Genes located in the upper-right quadrant show concordant positive co-dependencies across datasets, highlighting robust interaction candidates. **(C)** Network representation of first-degree neighbors of RBM48 and its top interactor ARMC7. Node color indicates direction of co-dependency, and edges represent significant interactions, illustrating local network structure and shared partners. **(D)** Summary table view of RBM48 co-dependencies within the HIDDENdb interface, showing standardized scores across datasets and supporting evidence from external resources.